\documentclass[aps,prb,twocolumn,groupedaddress]{revtex4}
\usepackage{makeidx}
\usepackage{amsmath,amssymb,amsfonts,amsthm}
\usepackage{graphicx,bm}

\begin{document}

\title{Anelastic spectroscopy study of the metal-insulator transition of Nd$%
_{1-x}$Eu$_{x}$NiO$_{3}$}
\date{}
\author{F. Cordero,$^{1}$ F. Trequattrini,$^{2}$ V.B. Barbeta,$^{3}$ R.F.
Jardim$^{4}$ and M.S. Torikachvili$^{5}$}
\affiliation{$^{1}$ CNR-ISC, Istituto dei Sistemi Complessi, Area della Ricerca di Roma -
Tor Vergata, Via del Fosso del Cavaliere 100, I-00133 Roma, Italy}
\affiliation{$^{2}$ Dipartimento di Fisica, Universit\`a di Roma \textquotedblleft La
Sapienza\textquotedblright , P.le A. Moro 2, I-00185 Roma, Italy}
\affiliation{$^{3}$ Departamento de F\'isica, Centro Universit\'ario da FEI, S\~ao
Bernardo do Campo, 09850-901, Brazil}
\affiliation{$^{4}$ Instituto de F\'isica, Universidade de S\~ao Paulo, CP 66318, S\~{a}o
Paulo, 05315-970, Brazil}
\affiliation{$^{5}$ Department of Physics, San Diego State University, San Diego,
California 92182-1233, USA}

\begin{abstract}
Measurements are presented of the complex dynamic Young's modulus of NdNiO$%
_{3}$ and Nd$_{0.65}$Eu$_{0.35}$NiO$_{3}$ through the Metal-Insulator
Transition (MIT). On cooling, the modulus presents a narrow dip at the MIT
followed by an abrupt stiffening of $\sim 6\%$. The anomaly is reproducible
between cooling and heating in Nd$_{0.65}$Eu$_{0.35}$NiO$_{3}$ but only
appears as a slow stiffening during cooling in undoped NdNiO$_{3}$,
conformingly with the fact that the MIT in $R$NiO$_{3}$ changes from
strongly first order to second order when the mean $R$ size is decreased.
The elastic anomaly seems not to be associated with the antiferromagnetic
transition, which is distinct from the MIT in Nd$_{0.65}$Eu$_{0.35}$NiO$_{3}$%
. It is concluded that the steplike stiffening is due to the disappearance
or freezing of dynamic Jahn-Teller (JT)\ distortions through the MIT, where
the JT active Ni$^{3+}$ is disproportionated into alternating Ni$^{3+\delta
} $ and Ni$^{3-\delta }$. The fluctuating octahedral JT distortion necessary
to justify the observed jump in the elastic modulus is estimated as $\sim
3\% $, but does not have a role in determining the MIT, since the otherwise
expected precursor softening is not observed.
\end{abstract}

\pacs{71.30.+h, 71.70.Ej, 62.40.+i, 75.50.Ee}
\maketitle

\section{Introduction}

There is still uncertainty on the exact nature of the metal-insulator
transition (MIT) in the $R$NiO$_{3}$ perovskites.\cite{Cat08,BSP11,LCB11}
Following a systematic study with several $R$ ion sizes,\cite{TLN92} the MIT
in these perovskites has been associated with the opening of a gap between
the O $2p$ band and the Ni $3d$ upper Hubbard band when the $R$ ion size
and/or temperature are decreased.\cite{TLN92,IFT98} In fact, the $R-$O bond
is too short with respect to the Ni$-$O one for the cubic perovskite
structure and the mismatch is relieved by tilting of the NiO$_{6}$
octahedra. Both a smaller $R$ size and cooling enhance the mismatch and
hence tilting, for steric reasons and because of the larger thermal
expansion of the $R-$O bond. This results in a further reduction of the
angle between Ni$-$O$-$Ni bonds with respect to 180$^{\mathrm{o}}$, reduces
the overlap and hence the width of the O $2p$ and Ni $3d$ bands and finally
opens a gap between them, causing the MIT.\cite{IFT98} Indeed, with
increasing the $R$ ion size the temperature $T_{\text{IM}}$ at which the MIT
occurs decreases, or the metallic phase becomes more stable.\cite{TLN92} In
this manner it is possible to rationalize the phase diagram of $R$NiO$_{3}$
of $T$ vs the $R$ ion size, and a similar behavior is found in cobaltites.%
\cite{TYK08}

On the other hand, Ni$^{3+}$ has an electronic configuration similar to that
of Mn$^{3+}$, with filled $t_{2g}$ triplet states (thanks to Hund's rule in
the case of Mn) and one electron in the $e_{g}$ doublet, whose degeneracy
can be lifted by tetragonal and orthorhombic distortions according to the
Jahn-Teller (JT) effect. The physics of Mn perovskites is dominated by the
Jahn-Teller coupling between these $e_{g}$ electronic states and the
distortions of the octahedra, which is a cause of orbital ordering (OO), and
the same might be expected for nickelates. On the contrary, according to the
early diffraction experiments the octahedral distortions in nickelates are
extremely small or null.\cite{Med97,IFT98} Later, starting with the
nickelates with smaller $R$, it has been found that the MIT is accompanied
by a subtle orthorhombic to monoclinic structural change, with charge
disproportionation (CD) 2Ni$^{3+}\rightarrow $ Ni$^{3+\delta }+$ Ni$%
^{3-\delta }$ and charge ordering (CO) into alternately expanded and
contracted NiO$_{6}$ octahedra along the three directions,\cite{AGF99} and
some octahedral distortion of the larger Ni$^{3-\delta }$O$_{6}$ octahedra.
A similar result has been recently found also with the larger $R$ ions Nd%
\cite{GAA09} and Dy.\cite{AMD08} Such JT distortions are found only in the
larger and more ionic Ni$^{3-\delta }$O$_{6}$ octahedra and are at least one
order of magnitude smaller than those observed in manganites.

Notice that in a naive picture CO and OO should not occur in the same set of
Ni ions, since Ni$^{3+}$ is JT active, while Ni$^{2+}$ and Ni$^{4+}$ are
not, but to what extent CO and OO occur in nickelates is still subject of
controversy. Just because of the subtleness of the structural changes
attributable to CO or OO, recourse has been had to resonant X-ray scattering.%
\cite{SMF02,LHP05,SSJ05,SSM06f} These experiments in NdNiO$_{3}$ are
interpreted as evidence of partial CO rather than OO occurring at the MIT.
According to this view, the degeneracy of the $e_{g}$ orbitals would be
lifted already in the metallic state, where the OO would be reflected by the
tilts of the octahedra; the MIT would be due only to CD, accompanied by a
small change of tilt angle.\cite{SSJ05} It is debated to what extent these
experiments may provide information on the contribution of CO and OO to MIT,%
\cite{DiM09} but the view that MIT is due to CO rather than OO received
further support from theory and experiments under hydrostatic pressure in
LuNiO$_{3}$,\cite{MKL07} and from the above mentioned diffraction studies on
$R=$ Nd and Eu, where alternated small and large octahedra are found.\cite%
{AMD08,GAA09} Nonetheless, recent experiments under pressure\cite{CZG10}
suggest that the previous ones\cite{MKL07} were affected by an unwanted
uniaxial component of pressure, and exclude a role of CO in the MIT;\cite%
{CZG10} rather, dynamic JT deformations would be responsible for large
isotope effects in $T_{\text{IM}}$.\cite{ZGD03}

Additional experiments that have been performed in order to probe CD with CO
and OO at the MIT in nickelates include M\"{o}ssbauer\cite%
{KDP02,PDB05,CML06,CMA07} and x-ray absorption spectroscopies.\cite{PTR05}
Again, there is no clear picture, with possibility of OO\cite{KDP02,PDB05}
or CO\cite{PDB05,CML06,CMA07} even in the metallic state.\cite{PTR05}

A similarly confused situation exists for NaNiO$_{2}$ and LiNiO$_{2}$, which
are also composed of NiO$_{6}$ octahedra but sharing the edges instead of
corners.\cite{KK05b} While NaNiO$_{2}$ undergoes a phase transformation with
clear cooperative JT distortion, LiNiO$_{2}$ has little or even reverse JT
distortion,\cite{CPS05} which is not yet explained. Some of the proposed
explanations involve the peculiar geometry of the LiNiO$_{2}$ lattice, where
the triangular Ni sublattice frustrates AFM correlations, but other possible
causes may apply also to $R$NiO$_{3}$, like the fact that the JT distortion
exists but is incoherent\cite{RDC95} or with very short coherence length,%
\cite{CPS05} or is hindered by the electron delocalisation.\cite{MKL07}

From the above summary and recent reviews\cite{Cat08} it appears that, in
spite of the wealth of experimental data on the $R$NiO$_{3}$ series, the
presence and role of CD and CO and/or OO at the MIT is not clear. In the Mn
perovskites important and quantitative information has been obtained on such
issues by studying the associated elastic anomalies;\cite{HGN00,HGN04} in Ni
perovskites no such type of investigation has ever been reported, except for
a preliminary report of the present results.\cite{BJT11} Here anelastic
spectroscopy measurements on Nd$_{1-x}$Eu$_{x}$NiO$_{3}$ are presented and
discussed, which shed new light on these issues.

\section{Experimental\label{sect Exp}}

Polycrystalline samples of Nd$_{1-x}$Eu$_{x}$NiO$_{3}$ ($x=0$ and 0.35) were
prepared from sol-gel precursors, sintered at $T=$ 1000~$^{\circ }$C, and
under oxygen pressure of 80~bar. Details of the synthesis process for
preparing these samples are described in details elsewhere.\cite{ESM00} All
samples were characterized by X-ray powder diffraction in a Brucker D8
Advance diffractometer. The x-ray diffraction patterns showed no extra
reflections due to impurity phases, and indicated that all samples have a
high degree of crystallinity. The samples were two bars with $x=0$ (NEN0)
and with $x=0.35$ (NEN35), both with approximate size $43\times 6\times 0.7$%
~mm$^{3}$ and a density $\rho \simeq 2.9(3.1)$~g/cm$^{3}$ for NEN0(35),
about $40\%\ $of the theoretical density. Such a high porosity makes the
evaluation of the absolute value of the complex Young's modulus $E=E^{\prime
}+iE^{\prime \prime }$ problematic, but did not affect the quality of the
measurement of its temperature dependence. This was measured by
electrostatically exciting the flexural vibrations of the bars, which were
suspended on thin thermocouple wires in vacuum, and with one face made well
conducting also below the MIT with Ag paint. It was possible to excite
various flexural modes, whose resonance frequencies are ideally in the
ratios $f_{\text{1F}}:$ $f_{\text{2F}}:f_{\text{3F}}:f_{\text{5F}}=$ $1:$ $%
2.800:$ $5.487:$ $13.55$. The fundamental frequency is given by\cite{NB72}
\begin{equation}
f_{\text{1F}}=1.028\frac{h}{l^{2}}\sqrt{\frac{E}{\rho }}  \label{fres}
\end{equation}%
where $l$ is the length, $h$ the thickness, $\rho $ the density of the bar;
at room temperature it was $f_{\text{1F}}\sim 0.6-1$~kHz.

In the Discussion we will need an estimate of the absolute value of the
Young's modulus, which is found with Eq. (\ref{fres}). The effective Young's
moduli at room temperature, uncorrected for porosity, were $E=4.5$~GPa for
NEN0 and 12.5~GPa for NEN35, which are $5-10$ times smaller than the typical
values $E\sim 100$~GPa found in ceramics of similar type. Sources of error
are the sample porosity, deviations from the shape of a rectangular
parallelepiped, low homogeneity and the presence of the Ag conductive layer,
wrong identification of the vibration modes and mixing of flexural with
torsional modes. The largest source of error in our case seems to be the
extremely high porosity, $p\simeq 0.60$. The other causes should be of minor
importance; for example, the NEN0 sample had the lowest value of $E$, but
its ratios $f_{\text{3F}}/f_{\text{1F}}=5.60$ and $f_{\text{5F}}/f_{\text{1F}%
}=13.66$ were in excellent agreement with the theoretical values given
above. The correction for such high porosities are very unreliable; to give
an idea, the Young's modulus of porous ceramics are corrected with empirical
expressions like $\left( 1-p\right) ^{n}$ with $n\simeq 1$,\cite{AKL01} or $%
\exp \left( -bp\right) $ with $2.1<$ $b<$ $2.8$.\cite{DYS02} These
expressions would enhance $E$ in our case by a factor between 2.5 and 5.2.
While this brings $E$ of NEN35 to reasonable values up to 66~GPa, the value
of NEN0 would remain below 25~GPa. The large difference between the two may
in part be accounted for by different types of porosities, in terms of shape
and connectivity of the pores, and in part due to the Eu substitution. In
the discussion we will assume $E\sim 65$~GPa.

The anelastic spectra are displayed in terms of the real part of the Young's
modulus or its reciprocal, the compliance $s=$ $E^{-1}=$ $s^{\prime
}-is^{\prime \prime }$, and of the elastic energy loss coefficient $Q^{-1}=$
$s^{\prime \prime }/s^{\prime }$.

\section{Results}

\begin{figure}[tbp]
\includegraphics[width=8.5 cm]{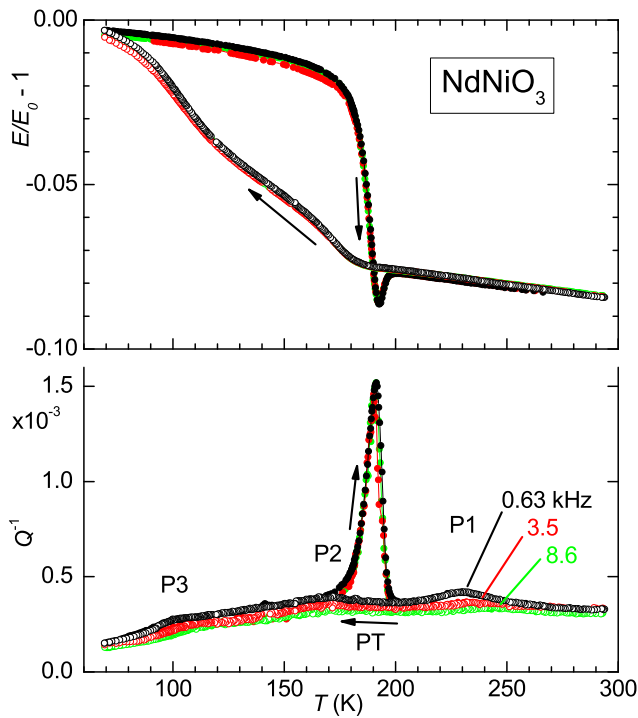} \vspace{0 cm}
\caption{Young's modulus $E$ and elastic energy loss coefficient $Q^{-1}$ of
NEN0 measured on cooling (empty symbols) and heating (filled symbols) at
0.63 kHz (black), 3.5 kHz (green)\ and 8.6 kHz (red) (colors online).}
\label{fig NEN0}
\end{figure}

In Fig. \ref{fig NEN0} is reported the anelastic spectrum of NEN0 measured
during cooling (empty symbols)\ and subsequent heating (filled symbols)
measured exciting the first odd flexural modes at 0.63, 3.5 and 8.6~kHz
during the same run. During cooling, below 190~K there is a progressive
stiffening of the Young's modulus in excess of the slight linear anharmonic
stiffening observed at higher temperature; such a stiffening proceeds until
the lowest temperature reached by us, although in the end it starts
levelling off. On heating, the modulus resumes the almost linear and weak
temperature dependence except for a steplike softening followed by a dip at
191~K. It is clear by comparison with resistivity, specific heat and
magnetization measurements that the steplike change of $E\left( T\right) $
with large temperature hysteresis is due to the MIT.\cite{EBJ06,BJT11}


The $Q^{-1}\left( T\right) $ is perfectly reproducible on heating and
cooling, except for the sharp peak in correspondence with the cusped
softening, both of which appear only during heating. In addition, there are
other three peaks: P1 at $230-240$~K, P2 at 171~K and P3 at $100-110$~K.
While the temperature of P2 is independent of frequency, the other two
increase with frequency, and therefore indicate thermally activated
processes whose maxima occur when the condition $\omega \tau \simeq 1$ is
met,\cite{NB72} where $\omega /2\pi $ is frequency and $\tau \left( T\right)
$ a relaxation time. We will ignore these reproducible but small peaks.

\begin{figure}[tbp]
\includegraphics[width=8.5 cm]{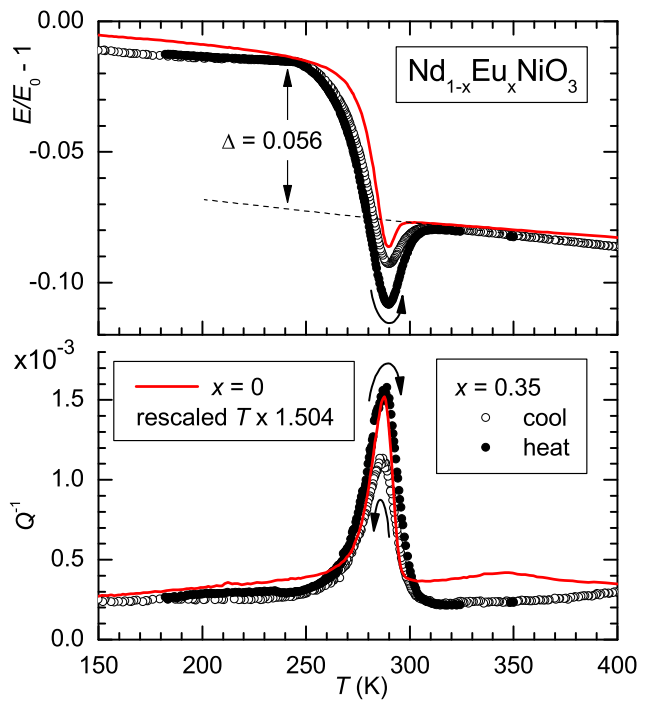} \vspace{0 cm}
\caption{Young's modulus $E$ and elastic energy loss coefficient $Q^{-1}$ of
NEN35 measured on cooling (empty symbols) and heating (filled symbols). For
comparison, the anelastic spectrum of NEN0 on heating is reported (line)
after scaling its tempearure to let its $T_{\text{IM}}$ coincide with that
of NEN35 (colors online).}
\label{fig NEN35}
\end{figure}

The spectrum of NEN35 (Fig. \ref{fig NEN35})\ is similar to that of NEN0
during heating, but the MIT occurs at $T_{\text{IM}}=$ 288~K, if identified
with the cusp in the real part, and the hysteresis is completely absent, the
only difference between heating and cooling being the intensity of the peak
in real and imaginary parts at the MIT. The curves measured at a frequency
five times higher are not shown for clarity, since they are identical except
that the peak in $Q^{-1}$ is $\sim 15\%$ higher, whereas in NEN0 it is
nearly identical.


The other small peaks in the $Q^{-1}$ are absent in NEN35. In Fig. \ref{fig
NEN35} are also reported the curves measured on NdNiO$_{3}$ during heating,
rescaled in temperature by a factor 1.504. The rescaled elastic anomaly of
NdNiO$_{3}$ is slightly sharper and narrower in the real part, with a
smaller dip, and almost coincident in the absorption. In NEN35 the
antiferromagnetic (AFM)\ transition occurs at $T_{\text{N}}=$ $240$~K $<T_{%
\text{IM}}$,\cite{BJT11} but there is no clear sign of it in the anelastic
spectra.

\section{Discussion}

The overall phenomenology of the anelastic anomalies found in Nd$_{1-x}$Eu$%
_{x}$NiO$_{3}$ appears to be mainly related to the MIT and in agreement with
what is known from other techniques. In NdNiO$_{3}$ the MIT occurs at $T_{%
\text{IM}}=T_{\text{N}}\simeq 192$~K and has marked first order character,
with a large hysteresis between cooling and heating. When decreasing the
mean $R$ ion size by Eu doping, the MIT shifts to higher temperature and
becomes second order, while $T_{\text{N}}$ shifts below $T_{\mathrm{IM}}$.
The anelastic anomalies also appear to be marginally affected by the
magnetic transition, since the heating curves are very similar to each other
regardless of the fact that $T_{\text{N}}$ coincides with $T_{\text{IM}}$ or
not. The present data can be compared with those of resistivity,
magnetization and specific heat taken on the same materials.\cite%
{EBJ06,BJT11} The hysteresis is due to the fraction of material that remains
metallic in the insulating phase, and the fact that the step in $\rho \left(
T\right) $ (see Fig. 7 in Ref. \onlinecite{EBJ06})\ is narrower than that in
$E\left( T\right) $ during cooling is explained by the fact that the latter
probes the bulk fraction of insulating and metallic phases, while
resistivity rather probes percolating conductive paths.

\subsection{Comparison with the specific heat anomaly}

\begin{figure}[tbp]
\includegraphics[width=8.5 cm]{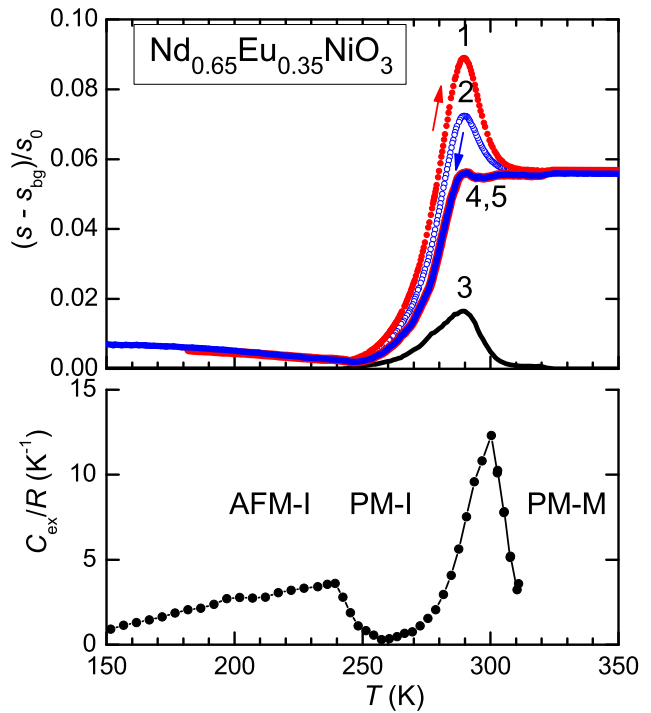} \vspace{0 cm}
\caption{(a)\ Curves 1 and 2: compliance $s=E^{-1}$ of NEN35 after
subtraction of the anharmonic background $s_{\text{bg}}$, measured on
heating and cooling. Curve 3 is the difference between them, $s_{\text{peak}}
$. Curve 4,5 are curves 1,2 after subtraction of the peaked component 3. (b)
Excess specific heat extracted from Ref. \onlinecite{BJT11}.}
\label{fig s_C}
\end{figure}

It seems that the dip in the modulus at $T_{\text{IM}}$ and the stiffening
below it have different origins, since with $x=0.35$ the dip and
accompanying peak in $Q^{-1}$ differ in amplitude between heating and
cooling, while the step in $E\ $is reproducible. This observation allows us
to separate the two anomalies and compare the dip in the modulus, or peak in
compliance $s=E^{-1}$, with the peak in specific heat. The curves 1 and 2 in
Fig. \ref{fig s_C} are the normalized compliance on heating and cooling,
after subtracting a linear background $s_{bg}$; the reference $s_0$ is the
extrapolation of $s_{bg}$ to 0~K. Besides the peak at 290~K and the step
below that temperature, there is an additional shallow rise of the
compliance below 250~K. The shape of the latter is very uncertain, due to
the background subtraction. The anharmonic phonon contribution to the
temperature dependence of the elastic constants is generally almost linear
and levels off at low temperature, but being unable to distinguish the
anharmonic contribution from the other anomalies in the temperature
dependence of $s$, we chose to use a simple linear interpolation over the
whole $60-350$~K range. Curve 3 is the difference between curves 1 and 2 and
it has been subtracted from them after multiplication by scale factors in
order to remove the peaked component. It turns out that the best values of
the factors are 2.00 and 1.00. The resulting curves 4 and 5 are coincident
and have only the step below 290~K plus the shallow rise.


Both the peak $s_{\text{peak}}$ (curve 3) and the excess compliance below
250~K can be put in relationship with the excess specific heat $C_{ex}$
measured on a similar sample.\cite{BJT11} In fact, a softening proportional
to the specific heat anomaly is expected at a magnetic transition,\cite%
{Lin79} and also at other types of transitions under the rather general
condition that their critical temperature depends on pressure or more
generally on stress.\cite{Tes75} A possible identification of the two
anomalies in $C_{ex}$ is the peak at $\sim 300$~K with the CO and/or orbital
transition concomitant with the MIT, possibly with participation of spin
degrees of freedom,\cite{KGG02} and the broad contribution below 250~K with
the onset of long range AFM order. The latter is expected to produce a jump $%
\Delta S=R\ln 2$ in the entropy of the $S=$ $\frac{1}{2}$ spins,\cite%
{BCG94,PBG99} but the jumps of the entropy $S=\int dT~C_{ex}/T$ are only $%
\Delta S=0.15R$ between $130$ and $260$~K and $\Delta S\simeq 0.12R$ between
$260$ and $320$~K. The fact that the actual step is smaller than expected
from complete AFM\ ordering has already been noted in other $R$NiO$_{3}$
nickelates, and has been interpreted as due to AFM correlations also above $%
T_{\text{N}}$, so that only a relatively small loss of entropy is involved
in the long range AFM ordering.\cite{BCG94,PBG99}

\subsection{Step-like stiffening below the MIT}

The main feature of the elastic anomaly at both compositions is the
stiffening below the MIT, whose step-like nature is made evident after the
subtraction of the peaked component in curves 4 and 5 of Fig. \ref{fig s_C}.
This is quite an unusual observation, since phase transitions are generally
accompanied by softening rather than stiffening. As discussed below, if some
strain component is coupled linearly with the order parameter of the
transition, then its elastic constant presents a cusped softening at the
critical temperature and is therefore followed by restiffening to the
background elastic constant on further cooling. This is not the case of
curves 4 and 5, which lack any sign of precursor softening divergent near $%
T_{\text{IM}}$.

There are several elastic studies of the CO and OO transitions in perovskite
manganites and in other compounds, but most of them concentrate on the
precursor softening above the transitions. In few cases the subsequent
restiffening below the transitions has been considered, especially if it
appears of excessive sharpness and amplitude with respect to the precursor
softening. In these cases recourse has been had to different sets of fitting
parameters above and below $T_{\text{CO ,}}$\cite{CZQ05} or to a Landau free
energy expansion with coefficients unrelated to CO or OO.\cite{SSS02} In Nd$%
_{0.5}$Sr$_{0.5}$MnO$_{3}$ a steplike stiffening at the MIT has been
interpreted as the effect of the renormalisation of the elastic constant by
the conduction electrons with a factor $\left( 1+g^{2}\chi \left( T\right)
\right) ^{-1/2}$, where $g$ is the electron-phonon coupling constant and $%
\chi \left( T\right) $ is the electron susceptibility, identified with the
magnetic susceptibility $\chi _{m}$.\cite{ZSL00} In that case, the MIT
coincides with a transition from FM to AFM and $\chi _{m}$ has a negative
step, which becomes positive in the renormalisation factor of the elastic
constant. This type of analysis is not appropriate to Nd$_{1-x}$Eu$_{x}$NiO$%
_{3}$, whose magnetic susceptibility has an almost imperceptible decrease
below $T_{\text{N}}$, superimposed to a stronger rise with cooling, even
after subtraction of the contribution of the Nd and Eu ions.\cite%
{ZGD00,EBJ06}

Lacking an adequate precedent for a satisfactory interpretation of the step
component of the elastic compliance of Nd$_{1-x}$Eu$_{x}$NiO$_{3}$, we first
review what kind of elastic response is expected at a MIT or magnetic
transition, and then propose an interpretation based on the response of the
JT distortions, from a point of view different from that usually adopted.
Detailed discussions of the elastic anomalies expected at various types of
phase transitions can be found in many review articles and books, usually
based on the Landau theory with the coupling between strain and order
parameter included,\cite{Reh73,BRN92,CS98,GL03,Lut07} and therefore we only
quote what is strictly necessary to our discussion.

\subsection{Landau analysis}

Near a phase transition the free energy can be expanded in powers of the
order parameter(s) $Q$. In the present case the order parameter (OP) can be
one or more of the symmetrized charge fluctuation coordinates for describing
charge ordering, or the quadrupolar orbital operators of the Ni ions for
orbital ordering, or the staggered magnetization for the AFM transition. By
including powers of $Q$\ up to the 6th order, it is possible to reproduce
both first and second order transitions, and including coupling terms
containing both $Q$ and strain $\varepsilon $, it is possible to deduce the
effect of the relaxation of the OP under stress on the elastic constants. In
the case that the OP is strain itself or is linearly coupled to it with a
term $\lambda \varepsilon Q$, the elastic constant coupled to the OP ideally
vanishes at the transition temperature $T_{\mathrm{C}}$, or at least has a
negative cusp. If bilinear coupling is forbidden because no strain has the
same symmetry of the OP, the next coupling term, $\mu \varepsilon Q^{2}$,
causes a negative step at $T_{\mathrm{C}}$ in the the elastic constant
coupled to the OP, possibly including a weak restiffening below $T_{\mathrm{C%
}}$. A biquadratic coupling $\nu \varepsilon ^{2}Q^{2}$ adds to the elastic
constant a contribution $2\nu \left\langle Q^{2}\right\rangle $ below $T_{%
\mathrm{C}}$, which can be a linear or saturating rise or decrease,
depending on the temperature dependence of the OP and on the sign of the
coupling constant $\nu $. Of all these terms, the latter is the only one
that might produce a stiffening below $T_{\mathrm{C}}$ without precursor
softening; other terms are possible but less important in the great majority
of cases.

\subsection{Magnetic transition}

Magnetic transitions cause elastic anomalies through essentially two
mechanisms: exchange striction and magnetostriction. The first mechanism
causes cusp-like softening in the elastic constants involving the strains
that change the atomic distances and hence the exchange constant. The
magnitude of such anomalies is of the order of 1\%.\cite{Lut07} Even smaller
step-like softening can be caused by magnetostrictive coupling with the
lattice, namely by a term $-B\varepsilon _{ij}S_{i}S_{j}$ linear in strain
and quadratic in spin variable or magnetization, which, according to
Landau's theory,\cite{Reh73} produces below the magnetic transition a
negative step in the modulus $M$ of magnitude $-2B^{2}/M$. Such a softening
is usually observed at magnetic transitions.\cite{Lin79,LWS69,KMB94}

On the other hand, in the FM transition of YTiO$_{3}$, a stiffening is found,%
\cite{MHI06} whose origin has not been explained, but also involves OO. In
the hexagonal quasi-1D AFM multiferroic YMnO$_{3}$, there is 1\% stiffening
below $T_{\text{N}}$ in $C_{11}$ and 3\% in $C_{66}$ due to biquadratic
coupling\cite{PLP07} (see previous paragraph). In fact, below $T_{\text{N}}=$
75~K the temperature dependence of $\Delta C_{ii}\left( T\right) /C_{ii}$
follows the squared OP, $S^{2}\propto $ $\left( 1-T/T_{\text{N}}\right)
^{2\beta }$, downto 2 K, which appears different from the abrupt rise in Nd$%
_{1-x}$Eu$_{x}$NiO$_{3}$. In the present case, the involvement of magnetic
effects in the steplike stiffening is made even more unlikely by the fact
that in Nd$_{0.65}$Eu$_{0.35}$NiO$_{3}$ the magnetic transition occurs at $%
T_{\text{N}}<T_{\text{IM}}$.

\subsection{Cooperative Jahn-Teller phase transition and orbital ordering
\label{sect JT}}

Following the notation of Hazama and coworkers,\cite{HGN00,HGN04} the
quadrupolar moments of the $e_{g}$ orbitals of Ni$^{3+}$ can couple with
tetragonal $\varepsilon ^{u}=\left( 2\varepsilon _{zz}-\varepsilon
_{xx}-\varepsilon _{yy}\right) /\sqrt{3}$ and orthorhombic $\varepsilon
^{v}=\varepsilon _{xx}-\varepsilon _{yy}$ strains (both with the same $E_{g}$
elastic constant $C^{\prime }=$ $\left( C_{11}-C_{12}\right) /2$), and
interact with each other causing the cooperative JT transition.\cite{Mel76}
The relevant part of the Hamiltonian of $N$ Ni$^{3+}$ ions per unit volume,
referred to the unit cell volume $v_{0}$, is
\begin{equation}
H=-v_{0}N\sum_{\gamma =u,v}g_{\gamma }O_{\gamma }\varepsilon ^{\gamma
}-v_{0}N\sum_{\gamma =u,v}g^{\prime }\left\langle O_{\gamma }\right\rangle
O_{\gamma ,i}  \label{H}
\end{equation}%
where the quadrupolar operators $O_{2}^{0}=\left(
2l_{z}^{2}-l_{x}^{2}-l_{y}^{2}\right) /\sqrt{3}$ and $%
O_{2}^{2}=l_{x}^{2}-l_{y}^{2}$ correspond, apart from a geometrical factor $%
a=\left\langle \phi _{eg}\right\vert O_{eg}\left\vert \phi
_{eg}\right\rangle =2\sqrt{3}$, to the occupation numbers of octahedra with
tetragonal and orthorhombic JT distortions reflecting the symmetry of their $%
3d$ orbitals. The first term is the linear coupling to the external strain,
with a coupling strength $g$, and causes softening of the $C^{\prime }$
elastic constant, while the second term is the elastic interaction among the
orbitals in the mean field approximation with strength $g^{\prime }$, and
determines the type of OO. The resulting softening of the $C^{\prime }$
elastic constant is
\begin{equation}
C^{\prime }=C_{0}^{\prime }-Ng^{2}\frac{\chi \left( T\right) }{1+g^{\prime
}\chi \left( T\right) }  \label{cJT1}
\end{equation}%
where $C_{0}^{\prime }$ is the background elastic constant and the
susceptibility
\begin{equation}
\chi \left( T\right) =\frac{a^{2}}{k_{\text{B}}T}  \label{chi}
\end{equation}%
is proportional to the contribution to the compliance $s=C^{-1}$ from
noninteracting orbitals, $\Delta s\left( T\right) =\left( \frac{g}{C^{\prime
}}\right) ^{2}\chi \left( T\right) $. Equation (\ref{cJT1}) can be rewritten
as
\begin{equation}
C^{\prime }=C_{0}^{\prime }\left( \frac{T-T_{\text{OO}}}{T-\Theta }\right)
~,\;  \label{cJT2}
\end{equation}%
where $\Theta =a^{2}g^{\prime }/k_{\text{B}}~,\;T_{\text{OO}}=\Theta +T_{%
\mathrm{JT}}$ and $T_{\mathrm{JT}}=Na^{2}g^{2}/\left( k_{\text{B}%
}C_{0}^{\prime }\right) $. In this form it is clear that the elastic
constant vanishes at $T_{\text{OO}}$, which is the onset of the OO\ or
cooperative JT transition, and that stiffens again on further cooling. At
higher temperatures, when $T\gg \Theta $ and hence $g^{\prime }\chi \left(
T\right) \ll 1$, Eq. (\ref{cJT1}) shows that the softening induced by the
orbitals freely responding to the applied stress is simply given by Eq. (\ref%
{chi}) and is $\propto $ $1/T$.

\subsection{Charge ordering}

The charge fluctuations can be expanded into fluctuation modes $Q_{\gamma }$
acting as OP of the CO transition, one or some of which may be linearly
coupled to the strain $\varepsilon _{\gamma }$ having the same symmetry with
terms $-g_{\gamma }Q_{\gamma }\varepsilon _{\gamma }$. As also shown in
Appendix B, if such terms are included in the expansion of the free energy
in powers of $Q_{\gamma }$ up to the fourth order, a second order transition
at a temperature $T_{\text{CO}}$ results, and the renormalized elastic
constant can be written exactly as in the case of the OO transition, Eq. (%
\ref{cJT2}), where $\Theta $ is the temperature at which the term $\frac{%
\alpha _{0}}{2}\left( T-\Theta \right) Q_{\gamma }^{2}$ of the Landau
expansion vanishes and $T_{\text{CO}}=\Theta +\Delta T$, $\Delta T=$ $%
g^{2}/\left( \alpha _{0}C_{\gamma }\right) $, with $C_{\gamma }$ the elastic
constant appropriate to $\varepsilon _{\gamma }$, is the temperature at
which the CO transition occurs with onset of spontaneous strain of type $%
\varepsilon _{\gamma }$. Therefore, OO\ between two orbitals and CO from
condensation of a single charge fluctuation mode produce the same precursor
softening on the appropriate elastic constant $C_{\gamma }$ ($=C^{\prime }$
for OO within the $e_{g}$ doublet). The restiffening below the transition is
discussed in Appendix B. Examples of materials similar to $R$NiO$_{3}$ where
both transitions have been recognized in ultrasonic experiments are La$%
_{1-x} $Sr$_{x}$MnO$_{3}$ with $T_{\text{CO}}<T_{\text{OO}}$\cite{HGN00} and
Pr$_{1-x}$Ca$_{x}$MnO$_{3}$ with $T_{\text{CO}}>T_{\text{OO}}$.\cite{HGN04}

\subsection{Softening in terms of octahedral JT\ distortion and its
disappearance below the MIT}

From the above discussion it appears that a transition whose driving force
is the interaction between JT active orbitals or between ionic charges would
appear in the coupled elastic constant as a clear precursor softening that
diverges at the transition, followed by restiffening on further cooling.
This is clearly not the case of Nd$_{1-x}$Eu$_{x}$NiO$_{3}$, since the
Young's modulus presents a weak and almost linear stiffening downto a
temperature very close to the MIT, which excludes any mechanism coupled to
strain. This is also shown quantitatively in Sect. I.

The picture we propose is the following. The MIT is an electronic
transition, presumably occurring at the opening of the charge
transfer gap during cooling,\cite{TLN92,IFT98} when the $R-$O bonds,
having larger thermal expansion of the shortest Ni$-$O bonds,
shorten faster than the latter and enhance the buckling of the
Ni$-$O$-$Ni bonds beyond a critical threshold. The driving mechanism
for this is the thermal contraction due to lattice anharmonicity,
and there is no effect on the elastic constants other than the usual
anharmonic stiffening. In the metallic phase the octahedra would be
JT distorted with a fluctuation rate much faster than our measuring
frequency, so that their contribution to the lattice softening would
be noncritical, namely Eq. (\ref{cJT1}) with $g^{\prime }=0$, and
the OO transition would occur well below $T_{\text{IM}}$.
\textit{{The MIT would inhibit such a softening, possibly due to the
concomitant CO that transforms JT active} Ni$^{3+}$ {into
(partially) inactive} Ni$^{3\pm \delta }$, {or making the
distortions static, or both. This would make the insulating phase
stiffer than the metallic one}}, so that the disappearance of the
metallic phase would be accompanied by elastic stiffening.

In what follows we will evaluate the JT distortion necessary to explain the
steplike stiffening in Figs. \ref{fig NEN0} and \ref{fig NEN35} with the
above mechanism and test the consistency of this explanation with the
available data. In order to relate the amount of elastic softening to the JT
distortion of the octahedra, rather than to the coupling coefficient $g$, it
is convenient to recast the above results in the usual formalism of
anelastic relaxation from point defects.\cite{NB72,33} It may seem
unrealistic to describe distorted octahedra sharing the O atoms, and hence
strongly interacting, as almost independent defects carrying a distortion,
but this is equivalent to the formulation of Sect. \ref{sect JT}, namely the
OO approach to the cooperative JT effect.\cite{KYB09} In the case of
nickelates it is certainly justified by the smallness of the JT distortions,
which make uncorrelated fluctuations less energetically unfavorable. If we
neglect the interaction term $\propto g^{\prime }$, the hamiltonian (\ref{H}%
) is equivalent to that of a system of independent defects of type $\gamma
=u,$ $v$ with double-force tensor $p^{\gamma }$, whose interaction energy
with an external strain $\varepsilon $ is\cite{LB78}
\begin{equation*}
E=-v_{0}\sum_{\gamma }\sum_{ij}c_{\gamma }p_{ij}^{\gamma }\varepsilon _{ji}
\end{equation*}%
where $c_{\gamma }$ with $\sum_{\gamma }c_{\gamma }=v_{0}N$ are the molar
concentrations of defects, and the summation over the cartesian components $%
i,j$ selects the strain component of the same symmetry of the defect, $%
\varepsilon ^{\gamma }$. It appears therefore that $Ng_{\gamma }O_{\gamma }$
in Eq. (\ref{H}) is equivalent to $c_{\gamma }p^{\gamma }$, although the
separate correspondences $NO_{\gamma }\rightarrow c_{\gamma }$ and $%
g_{\gamma }\rightarrow p^{\gamma }$ do not necessarily hold. In turn, the
double force tensor can be expressed in terms of the elastic quadrupole $%
\lambda _{ij}^{\gamma }$ (usually called elastic dipole\cite{NB72}) as $%
p_{ij}^{\gamma }=\sum_{kl}C_{ijkl}\lambda _{kl}^{\gamma }$, where $C$ is the
fourth-rank elastic stiffness tensor. The tensor $\lambda ^{\gamma }$ is
adimensional and is the strain of a unity concentration of defects of type $%
\gamma $, in the present case the tetragonal or orthorhombic strain of the
octahedra, so that the anelastic strain due to defects is
\begin{equation*}
\varepsilon ^{\mathrm{an}}=\sum_{\gamma }c_{\gamma }\lambda ^{\gamma },
\end{equation*}%
where the $c_{\gamma }$ obey Boltzmann's statistics and depend on
the applied stress $\sigma $ through the
defect energies $E_{\gamma }=$ $-v_{0}p^{\gamma }\varepsilon _{\gamma }=$ $%
-v_{0}\lambda ^{\gamma }\sigma _{\gamma }$. For only two types of defects
with the same symmetry and coupled to strain through the same symmetrized
elastic constant $C^{\prime }$ the relationship between $\lambda $ and $p$
is simply $\lambda =p/C^{\prime }$, and the contribution to the compliance, $%
s^{\prime }=1/C^{\prime }$, is\cite{NB72,33}
\begin{equation}
\delta s^{\prime }=\frac{d\varepsilon ^{\mathrm{an}}}{d\sigma }\propto c%
\frac{\left( \Delta \lambda \right) ^{2}}{k_{\text{B}}T}  \label{DS1}
\end{equation}%
where $\Delta \lambda =\left\vert \lambda _{u}-\lambda _{v}\right\vert $ and
the numerical factor differs from those usually reported for reorientation
of defects with the same symmetry, since we are dealing with the relaxation
between a tetragonal and an orthorhombic state rather the reorientation of
the principal axes of $\lambda $ among equivalent crystallographic
directions.

We can carry the analogy with point defects further into the dynamic
response. If $\tau $ is the relaxation time for an electron to change
orbital, the softening Eq. (\ref{DS1}) acquires a frequency dependent factor
$\left( 1+i\omega \tau \right) ^{-1}$ noticeable when temperature is lowered
enough to slow $\tau ^{-1}$ and make it comparable to the measuring angular
frequency $\omega $;\cite{NB72} this has also been discussed in connection
with JT effect in UO$_{2}$.\cite{BW68} The concentration $c$ of $e_{g}$
electrons would be 1 for $R$NiO$_{3}$ if there were no disproportionation of
Ni$^{3+}$. The MIT may reduce the magnitude of $\delta s^{\prime }$ both
reducing $c$ through the concomitant disproportionation of Ni$^{3+}$ and/or
making null the frequency factor if the distortions become quasistatic with $%
\omega \tau \gg 1$.

In Appendix A we define the octahedral distortions in terms of the
symmetrized tetragonal and orthorhombic strains $T^{u}$ and $T^{v}$
with magnitude $\lambda $, and evaluate the resulting
polycrystalline average of the relaxation strength of the Young's
modulus, $\Delta =$ $\delta E^{-1}/E^{-1}$, as

\textit{%
\begin{equation}
\left\langle \Delta \right\rangle \simeq \frac{cv_{0}\lambda ^{2}}{15k_{%
\text{B}}T}E~.  \label{relstr}
\end{equation}%
}

\subsection{Estimate of the Jahn-Teller distortion necessary to explain the
stiffening below the MIT}


As a first check that this mechanism can explain the stiffening at the MIT,
let us suppose that on passing across the MIT the JT\ distortion totally
disappears together with the Ni$^{3+}$ ions; therefore we equate the above $%
\left\langle \Delta \right\rangle $ with $c=1$ to the overall jump of $%
\delta s/s_{0}\simeq \delta E/E_{0}$. The largest source of error in
evaluating $\lambda $ is the absolute value of the Young's modulus, which we
assume to be $E\sim 65$~GPa (see Sect. \ref{sect Exp}), and together with $%
\Delta =0.056$ for NEN35 yields $\delta E^{-1}\simeq 10^{-13}$~cm$^{3}/$erg;
the cell volume is\cite{EBJ06} $v_{0}=55\times 10^{-24}~$cm$^{3}$ and Eq. (%
\ref{relstr}) with $T=T_{\mathrm{IM}}$ gives $\lambda \simeq 0.031$, namely
3\% octahedral and tetragonal distortions.

In order for this interpretation to be selfconsistent, it should turn out
that the temperature $T_{\text{OO}}=\Theta +T_{\mathrm{JT}}$ for OO\ is
smaller than $T_{\mathrm{IM}}$. There is no constraint to the smallness of
the intersite contribution $\Theta $, which can also be negative for
antiferroquadrupolar coupling. Instead, the JT energy is determined by the
coupling with uniform strain, $\lambda $, with $T_{\mathrm{JT}%
}=Na^{2}g^{2}/k_{\text{B}}C_{0}^{\prime }$, where $g$ is determined by the
equivalence $Nga\equiv cp=cC^{\prime }\lambda $. The $C_{0}^{\prime }$
elastic constant is unknown and we set it equal to $E$, finding $g/k_{\text{B%
}}\simeq 2300$~K and $T_{\mathrm{JT}}=240$~K, therefore smaller than $T_{%
\mathrm{IM}}$. The same estimate for NdNiO$_{3}$ with $\Delta =0.062$ and $%
T=190$~K yields $\lambda =0.026$, $g/k_{\text{B}}\simeq 1900$~K and $T_{%
\mathrm{JT}}=177$~K, again smaller than $T_{\mathrm{IM}}$.

Up to now it is shown that a CD with complete suppression of the JT effect
can explain the stiffening below the MIT, but it is also necessary to
determine the effect of an incomplete charge disproportionation 2Ni$%
^{3+}\rightarrow $ Ni$^{3+\delta }+$ Ni$^{3-\delta }$ with $\delta <1$. The
above treatment suggests a naive analogy of JT active octahedra with point
defects, whose concentration is $c=1$ in the conductive phase and becomes $%
c=1-\delta $ in the insulating one, with unchanged distortion $\lambda $.
However, this is not the picture proposed after the diffraction experiments,
where $\delta $ is deduced from the change of valence of Ni$^{3+}$, and this
is in turn extracted from the average Ni$-$O bond lengths together with the
bond valence model.\cite{GAA09} In this picture, all the octahedra undergo
disproportionation, and are alternated along the three principal directions
as small Ni$^{3+\delta }$ and large Ni$^{3-\delta }$ octahedra, only the
latter supporting some residual JT distortion.\cite{GAA09,AMD08} From this
point of view, a more sensible hypothesis is that, when entering the
insulating phase, in Eq. (\ref{relstr}) $c$ passes from 1 to $1/2$ and the
JT distortion $\lambda $ is reduced by an amount depending on $\delta $. If
we suppose a linear relationship between $\lambda $ and $\delta $, then we
should set $\lambda \rightarrow \lambda \left( 1-\delta \right) $ into Eq. (%
\ref{relstr}), and repeating the above analysis with the estimate $\delta
\simeq 0.28$ for NdNiO$_{3}$,\cite{GAA09,SMF02} and $c=1/2$ we would find $%
\lambda =0.036$ and $T_{\mathrm{JT}}=330$~K. The latter value, being
slightly larger than $T_{\mathrm{IM}}$, seems at first inconsistent with the
present analysis, but there are several arguments that this is not
necessarily the case. A\ first point is that we ignored the correlations
between octahedra, which in the perovskite structure with octahedra sharing
the vertices is expected to be antiferroelastic\cite{KK82} and corresponds
to a negative $\Theta $ compensating $T_{\mathrm{JT}}$, as in Pr$_{1-x}$Ca$%
_{x}$MnO$_{3}$.\cite{HGN04} Another point is that the JT distortion
may not fall linearly with $\delta $ but faster. Finally, if the JT
distortions remained frozen below the MIT, then the kinetic factor
$1/\left[ 1+\left( \omega \tau \right) ^{2}\right] $ would pass from
1 to 0 on crossing the MIT and the initial safe estimate of $\lambda
$ of Eq.(\ref{relstr}) with $c=1$ would again be correct.

Once established that the stiffening below the MIT can be consistently
explained by vanishing or blocking of dynamic JT distortions $\lambda $, it
is necessary to check the compatibility of such disordered distortions above
$T_{\mathrm{IM}}$ with the structural studies on NdNiO$_{3}$. We refer to
the most recent experiment with synchrotron powder-diffraction,\cite{GAA09}
where the Ni$-$O\ bond lengths are reported to be $l\simeq 1.9$~\AA , and
the B-factors of the apical and equatorial O atoms are 0.72 and 1.13~\AA $%
^{2}$ at room temperature, resulting in thermal plus disordered
displacements of 0.1 and 0.12~\AA\ respectively. The JT distortions
estimated here of $\lambda \sim 0.03$ require changes of the bond lengths of
$\delta l=\lambda \times l\sim 0.06$~\AA , which, being dynamically
disordered, would appear only in the Debye-Waller factors, and are about
half of the B-factor displacements at room temperature in NdNiO$_{3}$. The
JT distortions required to explain the stiffening at the MIT are therefore
compatible with the diffraction experiments:\ above the MIT they are
disordered and account for half of the thermal/disordered displacements of
the O\ atoms; below the MIT they are totally or partially absent due to the
disproportionation of the JT Ni$^{3+}$ into Ni$^{3\pm \delta }$, or simply
frozen. This may also explain EXAFS experiments indicating that the
splitting of the Ni$-$O bond lengths persist also above the MIT in $R$NiO$%
_{3}$ with $R=$ Pr, Nd, Eu.\cite{PTR05} This fact has been explained in
terms of localized CO phase also in the metallic state, but might be due to
the disordered JT distortions. Another case where the anomalously large
mean-square displacements suggest JT distortions is a recent neutron
diffraction study on PrNiO$_{3}$.\cite{MFL08}

\subsection{The peaked softening at $T_{\text{IM}}$}

Up to now we dealt with the steplike stiffening below the MIT, which
however is accompanied by a peaked softening and absorption. The
narrow shape of such an anomaly suggests that it is due to
fluctuations associated with the MIT, but the absorption is almost
independent of frequency, while the critical absorption is expected
to be proportional to it.\cite{Lut07}


\begin{figure}[tbp]
\textit{\includegraphics[width=8.5 cm]{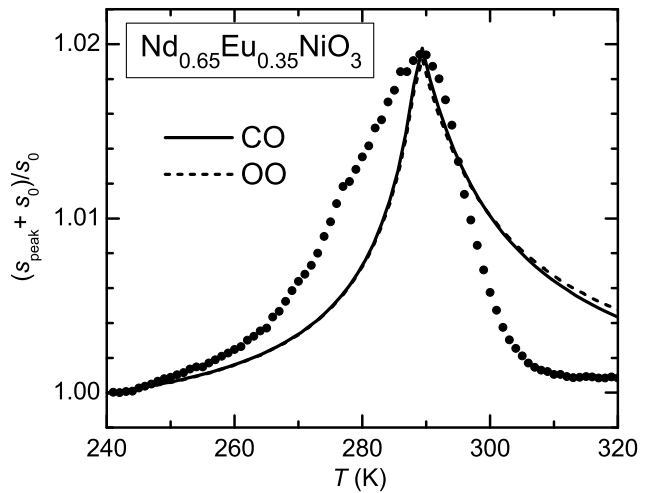} \vspace{0 cm} }
\caption{Peaked component of $s^{\prime }$ (curve 3 of Fig. \protect\ref{fig
s_C}) fitted assuming CO or OO at 289~K.}
\label{fig fitpeak}
\end{figure}

The real part of the peak, corresponding to curve 3 in Fig. \ref{fig
s_C}, is plotted in Fig. \ref{fig fitpeak} as $s_{\text{peak}} +
s_0$, cleared of the step and temperature dependent part of the
background. We tried to fit with the expressions appropriate for CO
and OO, namely $s =$ $1/\left(C_{\text{CO/OO}} +
C_{\text{bg}}\right)$ where $C_{\text{CO/OO}}\left( T\right) $ is
given by Eq. (\ref {cJT2}) above $T_{\text{CO}}/T_{\text{OO}}$ and
in Appendix B in the low
temperature phase. The parameters were $T_{\text{CO}}=T_{\text{OO}}=289$~K, $%
\Delta T=$ $T_{\mathrm{JT}}=$ $14.4$~K and $C_{0}\simeq 0.02$,
$C_{\text{bg}} = s_0^{-1} = 0.98$ for both cases. The resulting
curves, very close to each other, fit very poorly the experimental
data. In particular, the observed peak is steeper in the high
temperature side, while the fitting curves are steeper in the low
temperature phase, as expected. This discrepancy cannot be accounted
for by relaxation of domain walls in the low temperature phase,
since relaxations depend on the measuring frequency and the anomaly
we observe does not. The failure of reproducing the peaked component
as softening from OO or CO should be considered as an indication
that the origin of the peak is different and we leave it as an open
question.

\section{Conclusions}

We measured the complex dynamic Young's modulus of NdNiO$_{3}$ and Nd$_{0.65}
$Eu$_{0.35}$NiO$_{3}$, in order to study the elastic anomalies associated
with the MIT and magnetic transition. The main features are a sharp
stiffening of $\sim 6\%$ below the MIT, perfectly reproducible during
cooling and heating in Nd$_{0.65}$Eu$_{0.35}$NiO$_{3}$ and with a broad
hysteresis in NdNiO$_{3}$, accompanied by a narrow dip at the MIT. The
change of the MIT between first- and second-order character at the two
compositions agrees with the known phase diagram of $R$NiO$_{3}$ as a
function of the $R$ size. The differences with respect to elasticity
measurements on Mn perovskites are the absence of precursor softening
associated with the MIT, the sharpness of the transition during heating or
when it is second order, and the broadness of the hysteresis when it is
first order. Both the narrow softening at the MIT and the stiffening below
it seem not to be associated with spin ordering, which in the Eu-doped
sample occurs at a temperature lower than the MIT.

By comparing cooling and heating runs, it was possible to separate the
steplike stiffening from the other anomalies, which present similarities
with the excess specific heat, and whose relationship with spin and charge
order have been briefly considered. A fit of the dip at the MIT in terms of
CO has been made, but its quality is too poor to draw a conclusion.

In order to explain the rather unusual steplike stiffening, possible
mechanisms producing elastic anomalies at magnetic, charge order and orbital
order transitions have been considered, and it is concluded that the
stiffening is due to the disappearance or to freezing in the insulating
phase of dynamic JT\ distortions, due to the charge disproportionation and
ordering that accompanies the MIT and transforms JT active Ni$^{3+}$ into Ni$%
^{3\pm \delta }$. The driving force for the MIT is not orbital ordering or
any mechanism linearly coupled to strain, in view of the absence of
precursor softening. This is in agreement with the idea that the MIT\ is
caused by the opening of a gap when the orbital overlap drops below a
threshold during cooling. In fact, such a mechanism depends only on the
different thermal expansivities of the $R-$O and Ni$-$O bonds, and does not
produce softening. The fluctuating tetragonal/orthorhombic Jahn-Teller
distortion of the NiO$_{6}$ octahedra necessary to justify the observed jump
in the elastic modulus is estimated as $\sim 3\%$ and the compatibility of
such a requirement with other structural studies is discussed.

\subsection*{Acknowledgments}

The authors thank Prof. Marcia T. Escote and Fenelon M. Pontes for the
preparation of the samples. F.C. wishes to thank Mr. P.M. Latino and F.
Corvasce for their technical assistance in the experiments. This work was
partially supported by the Brazil agencies FAPESP and CNPq.

\subsection{Appendix A}

We determine the factor that connects the amplitude of softening in Eq. (\ref%
{DS1}) to the anisotropy of the elastic quadrupole tensor of the octahedral
distortions $\Delta \lambda $. This can be done by evaluating the relaxation
of the reciprocal of the Young's modulus, $\delta E^{-1}\left( \mathbf{\hat{n%
}}\right) $, along a generic direction $\hat{n}$ after application of a
uniaxial stress $\sigma =\sigma _{0}\left\vert \mathbf{\hat{n}}\right\rangle
\left\langle \mathbf{\hat{n}}\right\vert $ and then performing the angular
average of the relaxation strength, $\delta E^{-1}\left( \mathbf{\hat{n}}%
\right) /E^{-1}\left( \mathbf{\hat{n}}\right) $, over $\hat{n}$, in order to
obtain an approximation of the polycrystalline average. It is convenient to
decompose stress and strain into an orthonormal basis adapted to the cubic
symmetry\cite{LB78}
\begin{eqnarray}
T^{1} &=&\frac{1}{\sqrt{3}}\left[
\begin{array}{ccc}
1 & 0 & 0 \\
0 & 1 & 0 \\
0 & 0 & 1%
\end{array}%
\right] ,  \notag \\
T^{u} &=&\frac{1}{\sqrt{6}}\left[
\begin{array}{ccc}
-1 & 0 & 0 \\
0 & -1 & 0 \\
0 & 0 & 2%
\end{array}%
\right] ~,\;T^{v}=\frac{1}{\sqrt{2}}\left[
\begin{array}{ccc}
1 & 0 & 0 \\
0 & -1 & 0 \\
0 & 0 & 0%
\end{array}%
\right] ~  \label{basis}
\end{eqnarray}%
where $T^{1}$ is the dilatation (irreducible representation $A_{1g}$), $T^{u}
$ and $T^{v}$ tetragonal and orthorhombic shears of type $E_{g}$, and the
three shears of type $T_{2g}$ do not enter in the calculation, since they
are orthogonal to the $E_{g}$ JT distortion. The generic strain $\varepsilon
$ can be decomposed as $\left\vert \varepsilon \right\rangle =$ $%
\sum_{\gamma }\varepsilon _{\gamma }\left\vert T^{\gamma }\right\rangle $,
or explicitly $\varepsilon _{ij}=\sum_{\gamma }\varepsilon _{\gamma
}T_{ij}^{\gamma }$ with components $\varepsilon _{\gamma }=\left\langle
T^{\gamma }\right\vert \left. \varepsilon \right\rangle $, the scalar
product being defined as Tr$\left\{ T^{\gamma }:\varepsilon \right\}
=\sum_{ij}T_{ij}^{\gamma }\varepsilon _{ji}$. Analogously, the uniaxial
stress along $\hat{n}$ is decomposed as
\begin{equation*}
\sigma _{ij}=\sigma _{0}n_{i}n_{j}=\sum_{\gamma }\sigma _{\gamma
}T_{ij}^{\gamma }~.
\end{equation*}%
Thanks to the orthonormality of the $T^{\gamma }$ only the two components $%
\sigma _{u}$ and $\sigma _{v}$ will couple with $\lambda ^{u}$ and $\lambda
^{v}$ and we write them as
\begin{equation*}
\sigma _{u}=\sigma _{0}u,\;\sigma _{v}=\sigma _{0}v
\end{equation*}%
with%
\begin{eqnarray}
u &=&\left\langle \mathbf{\hat{n}}\right\vert T^{u}\left\vert \mathbf{\hat{n}%
}\right\rangle =\frac{2n_{33}^{2}-n_{11}^{2}-n_{22}^{2}}{\sqrt{6}}
\label{uv} \\
v &=&\left\langle \mathbf{\hat{n}}\right\vert T^{v}\left\vert \mathbf{\hat{n}%
}\right\rangle =\frac{n_{11}^{2}-n_{22}^{2}}{\sqrt{2}}~.  \notag
\end{eqnarray}%
The elastic quadrupoles of the two JT states are written as
\begin{equation*}
\lambda ^{u}=\lambda T^{u}~,\;\lambda ^{v}=\lambda T^{v}
\end{equation*}%
where $\lambda $ gives the strength of the distortion.

The application of $\sigma $ changes their elastic energies by $d\left(
E_{u}-E_{v}\right) =$ $-v_{0}\sum_{ij}\left( \lambda _{ij}^{u}-\lambda
_{ij}^{v}\right) \sigma _{ji}=$ $-v_{0}\sigma _{0}\lambda \left( u-v\right) $
and their populations,
\begin{equation*}
n_{u}=e^{-E_{u}/k_{\text{B}}T}/\left( e^{-E_{u}/k_{\text{B}}T}+e^{-E_{v}/k_{%
\text{B}}T}\right)
\end{equation*}%
and $n_{v}=1-n_{u}$ change as $\frac{dn_{u}}{d\sigma _{0}}\simeq \frac{v_{0}%
}{4k_{\text{B}}T}\lambda \left( v-u\right) $, where it is used $\lambda
\sigma \ll k_{\text{B}}T$. This change of populations results in a change of
the anelastic strain
\begin{eqnarray*}
d\varepsilon ^{\text{an}} &=&c\sum_{\gamma }\lambda ^{\gamma }dn_{\gamma }=
\\
&=&\frac{\sigma _{0}v_{0}c}{4k_{\text{B}}T}\lambda ^{2}\left( u-v\right) %
\left[ T^{v}-T^{u}\right]
\end{eqnarray*}%
. The uniaxial component of $d\varepsilon ^{\text{an}}$ parallel to $\hat{n}$
is $\left\langle \mathbf{\hat{n}}\right\vert d\varepsilon ^{\text{an}%
}\left\vert \mathbf{\hat{n}}\right\rangle =$ $\sum_{ij}n_{i}n_{j}d%
\varepsilon _{ij}^{\text{an}}$, and using Eq. (\ref{uv}) or $\left\langle
\mathbf{\hat{n}}\right\vert T^{v}-T^{u}\left\vert \mathbf{\hat{n}}%
\right\rangle =$ $\left( u-v\right) $, the relaxation of the Young's modulus
along $\hat{n}$ is
\begin{equation*}
\delta E^{-1}\left( \mathbf{\hat{n}}\right) =\frac{\left\langle \mathbf{\hat{%
n}}\right\vert d\varepsilon ^{\text{an}}\left\vert \mathbf{\hat{n}}%
\right\rangle }{\sigma _{0}}=\frac{cv_{0}}{4k_{\text{B}}T}\lambda ^{2}\left(
u-v\right) ^{2}
\end{equation*}%
A simple formula for the polycrystalline average of the relaxation strength $%
\Delta \left( \mathbf{\hat{n}}\right) =$ $\delta E^{-1}\left( \mathbf{\hat{n}%
}\right) /E^{-1}\left( \mathbf{\hat{n}}\right) $ is obtained from the Reuss
approximation of uniform stress over the grains,\cite{NB72} taking the
angular average $\left\langle \Delta \right\rangle \simeq $ $\left\langle
\delta E^{-1}\left( \mathbf{\hat{n}}\right) \right\rangle E$ where the
anisotropy of the material is neglected. By using Eq. (\ref{uv}) and $%
\left\langle n_{i}^{2}\right\rangle =\frac{1}{3}$, $\left\langle
n_{i}^{4}\right\rangle =\frac{1}{5}$, $\left\langle
n_{i}^{2}n_{j}^{2}\right\rangle =\frac{1}{15}$ one finally obtains
\begin{equation*}
\left\langle \Delta \right\rangle \simeq \frac{cv_{0}\lambda ^{2}}{15k_{%
\text{B}}T}E~.
\end{equation*}

It is possible to introduce a mixing of the two normalized symmetry strains,
due to hybridisation with the O$2p$ orbitals and delocalisation,\cite{BA04}
by setting $\lambda ^{u}=\lambda \left[ \left( 1-\frac{f}{2}\right) T^{u}+%
\frac{f}{2}T^{v}\right] $ and similarly for $\lambda ^{v}$ with $0\leq f\leq
1$. In this case, in the above formulas $\lambda $ should be replaced by $%
\lambda \left( 1-f\right) $, but from anelastic measurements it is not
possible to evaluate $\lambda $ and $f$ independently, and we assume $f=0$.

\subsection{Appendix B}

The softening in Eq. (\ref{cJT2}) is valid above the CO or OO
transition temperature; below that temperature, the restiffening
depends on the precise nature of the transition. In the simplest
cases of doublet OO,\cite{Mel76} the susceptibility below
$T_{\mathrm{C}}$ is given by
\begin{equation*}
\chi =\frac{1}{k_{\text{B}}T\cosh ^{2}\left( h/k_{\text{B}}T\right) }
\end{equation*}%
where $h$ is the mean field felt by each ion, acting as order parameter, and
is solution of the self-consistent equation
\begin{equation*}
h=k_{\text{B}}T_{\mathrm{C}}\tanh \left( h/k_{\text{B}}T\right) ~.
\end{equation*}

For CO, consider the Landau expansion in terms of powers of the charge
fluctuation order parameter, as in Ref. [\onlinecite{GL03}] but limiting to
the case that only one order parameter $Q$ is relevant. Than the free energy
is written as
\begin{equation*}
F=F_{0}+\frac{\alpha }{2}Q^{2}+\frac{\beta }{4}Q^{4}+\frac{C_{0}}{2}%
\varepsilon ^{2}-gQ\varepsilon
\end{equation*}%
where $\alpha =\alpha _{0}\left( T-\Theta \right) $. Solving the equilibrium
conditions $\partial F/\partial Q=0$ and $\partial F/\partial Q\varepsilon =0
$ one finds $Q^{2}\left( T\right) =\alpha _{0}/\beta ~\left( T-T_{\text{CO}%
}\right) $ with $T_{\text{CO}}=\Theta +\Delta T$, $\Delta T=g^{2}/\left(
C_{0}\alpha _{0}\right) $. The renormalized elastic constant is then
\begin{equation*}
C=C_{0}-g\frac{\partial Q}{\partial \varepsilon }
\end{equation*}%
where $\partial Q/\partial \varepsilon $ can be obtained by derivating with
respect to $\varepsilon $ the equation $\partial F/\partial Q=0$:
\begin{equation*}
\frac{\partial Q}{\partial \varepsilon }=\frac{g}{\alpha +3\beta Q^{2}}~.
\end{equation*}%
When $T>\Theta $ it is $Q=0$ and we obtain Eq. (\ref{cJT2}) as before, here
rewritten as
\begin{equation*}
C=C_{0}\frac{\left( T-\Theta -\Delta T\right) }{\left( T-\Theta \right) }
\end{equation*}%
In the CO phase, we obtain
\begin{equation*}
C=C_{0}\frac{\left( T-\Theta -\Delta T\right) }{\left( T-\Theta -\frac{3}{2}%
\Delta T\right) }~.
\end{equation*}

These formulas have been used for the fits in Fig. \ref{fig
fitpeak}.


\begin{references}

\bibitem{Cat08}
G. Catalan, Phase Trans. {\bf 81}, 729 (2008).

\bibitem{BSP11}
Y. Bodenthin, U. Staub, C. Piamonteze, M.
Garc{\'{\i}}a-Fern{\'a}ndez, M.J. Mart{\'{\i}}nez-Lope and J.A.
Alonso, J. Phys.: Condens. Matter {\bf 23}, 036002 (2011).

\bibitem{LCB11}
S. B. Lee, R. Chen and L. Balents, Phys. Rev. Lett. {\bf 106}, 016405
(2011).

\bibitem{TLN92}
J.B. Torrance, P. Lacorre, A.I. Nazzal, E.J. Ansaldo and Ch. Niedermayer,
Phys. Rev. B {\bf 45}, 8209 (1992).

\bibitem{IFT98}
M. Imada, A. Fujimori and Y. Tokura, Rev. Mod. Phys. {\bf 70}, 1039
(1998).

\bibitem{TYK08}
M. Tachibana, T. Yoshida, H. Kawaji, T. Atake and E.
Takayama-Muromachi, Phys. Rev. B {\bf 77}, 094402 (2008).

\bibitem{Med97}
M.L. Medarde, J. Phys.: Condens. Matter {\bf 9}, 1679 (1997).

\bibitem{AGF99}
J.A. Alonso, J.L. Garc{\'{\i}}a-Mu{\~n}oz, M.T.
Fern{\'a}ndez-D{\'{\i}}az, M.A.G. Aranda, M.J. Mart{\'{\i}}nez-Lope
and M.T. Casais, Phys. Rev. Lett. {\bf 82}, 3871 (1999).

\bibitem{GAA09}
J.L. Garc\'ia-Mu{\~n}oz, M.A.G. Aranda, J.A. Alonso and M.J.
Mart\'inez-Lope, Phys. Rev. B {\bf 79}, 134432 (2009).

\bibitem{AMD08}
J.A. Alonso, M.J. Mart{\'{\i}}nez-Lope, G. Demazeau, M.T.
Fern{\'a}ndez- D{\'{\i}}az, I.A. Presniakov, V.S. Rusakov, T.V.
Gubaidulina and A.V. Sobolev, Dalton Trans. {\bf2008}, 6584.

\bibitem{SMF02}
U. Staub, G.I. Meijer, F. Fauth, R. Allenspach, J.G. Bednorz, J.
Karpinski, S.M. Kazakov, L. Paolasini and F. d'Acapito, Phys. Rev.
Lett. {\bf 88}, 126402 (2002).

\bibitem{LHP05}
J.E. Lorenzo, J.L. Hodeau, L. Paolasini, S. Lefloch, J.A. Alonso and
G. Demazeau, Phys. Rev. B {\bf 71}, 045128 (2005).

\bibitem{SSJ05}
V. Scagnoli, U. Staub, M. Janousch, A.M. Mulders, M. Shi, G.I.
Meijer, S. Rosenkranz, S.B. Wilkins, L. Paolasini, J. Karpinski,
S.M. Kazakov and S.W. Lovesey, Phys. Rev. B {\bf 72}, 155111 (2005).

\bibitem{SSM06f}
V. Scagnoli, U. Staub, A.M. Mulders, M. Janousch, G.I. Meijer, G.
Hammerl, J.M. Tonnerre and N. Stojic, Phys. Rev. B {\bf 73}, 100409
(2006).

\bibitem{DiM09}
S. Di Matteo, J. Phys.: Conf. Series {\bf 190}, 012008 (2009).

\bibitem{MKL07}
I.I. Mazin, D.I. Khomskii, R. Lengsdorf, J.A. Alonso, W.G. Marshall,
R.M. Ibberson, A. Podlesnyak, M.J. Mart{\'{\i}}nez-Lope and M.M.
Abd-Elmeguid, Phys. Rev. Lett. {\bf 98}, 176406 (2007).

\bibitem{CZG10}
J.-G. Cheng, J.-S. Zhou, J.B. Goodenough, J.A. Alonso and M.J.
Martinez-Lope, Phys. Rev. B {\bf 82}, 085107 (2010).

\bibitem{ZGD03}
J.-S. Zhou, J.B. Goodenough and B. Dabrowski, Phys. Rev. B {\bf 67},
020404 (2003).

\bibitem{KDP02}
S.-J. Kim, G. Demazeau, I. Presniakov, K. Pokholok, A. Baranov, A.
Sobolev, D. Pankratov and N. Ovanesyan, Phys. Rev. B {\bf 66},
014427 (2002).

\bibitem{PDB05}
I. Presniakov, G. Demazeau, A. Baranov, A. Sobolev and K. Pokholok,
Phys. Rev. B {\bf 71}, 054409 (2005).

\bibitem{CML06}
A. Caytuero, H. Micklitz, F.J. Litterst, E.M. Baggio-Saitovitch,
M.M. Abd-Elmeguid and J.A. Alonso, Phys. Rev. B {\bf 74}, 094433
(2006).

\bibitem{CMA07}
A. Caytuero, H. Micklitz, M.M. Abd-Elmeguid, F.J. Litterst, J.A.
Alonso and E.M. Baggio-Saitovitch, Phys. Rev. B {\bf 76}, 193105
(2007).

\bibitem{PTR05}
C. Piamonteze, H.C.N. Tolentino, A.Y. Ramos, N.E. Massa, J.A. Alonso,
M. J. Mart{\'{\i}}nez-Lope and M.T. Casais, Phys. Rev. B {\bf
71}, 012104 (2005).

\bibitem{KK05b}
P. Kalyani and N. Kalaiselv, Sci. Technol. Adv. Mater. {\bf 6}, 689
(2005).

\bibitem{CPS05}
J.-H. Chung, Th. Proffen, S. Shamoto, A.M. Ghorayeb, L. Croguennec,
W. Tian, B.C. Sales, R. Jin, D. Mandrus and T. Egami, Phys. Rev. B
{\bf 71}, 064410 (2005).

\bibitem{RDC95}
A. Rougier, C. Delmas and A.V. Chadwick, Solid State Commun. {\bf
94}, 123 (1995).

\bibitem{HGN00}
H. Hazama, T. Goto, Y. Nemoto, Y. Tomioka, A. Asamitsu and Y.
Tokura, Phys. Rev. B {\bf 62}, 15012 (2000).

\bibitem{HGN04}
H. Hazama, T. Goto, Y. Nemoto, Y. Tomioka, A. Asamitsu and Y.
Tokura, Phys. Rev. B {\bf 69}, 064406 (2004).

\bibitem{BJT11}
V.B. Barbeta, R.F. Jardim, M.S. Torikachvili, M.T. Escote, F.
Cordero, F.M. Pontes and F. Trequattrini, J. Appl. Phys. {\bf 109},
07E115 (2011).

\bibitem{ESM00}
M.T. Escote, A.M.L. da Silva, J.R. Matos and R.F. Jardim
J. Solid State Chem.  {\bf 151}, 298 (2000).

\bibitem{NB72}
A.S. Nowick and B.S. Berry, {\it Anelastic Relaxation in Crystalline
Solids}. (Academic Press, New York, 1972).

\bibitem{AKL01}
M. Asmani, C. Kermel, A. Leriche and M. Ourak, J. Eur. Ceram. Soc.
{\bf 21}, 1081 (2001).

\bibitem{DYS02}
R.A. Dorey, J.A. Yeomans and P.A. Smith, J. Eur. Ceram. Soc. {\bf
22}, 403 (2002).

\bibitem{EBJ06}
M.T. Escote, V.B. Barbeta, R.F. Jardim and J. Campo, J. Phys.:
Condens. Matter {\bf 18}, 6117 (2006).

\bibitem{Lin79}
M.E. Lines, Phys. Rep. {\bf 55}, 133 (1979).

\bibitem{Tes75}
L.R. Testardi, Phys. Rev. B {\bf 12}, 3849 (1975).

\bibitem{KGG02}
R. Klingeler, J. Geck, R. Gross, L. Pinsard-Gaudart, A.
Revcolevschi, S. Uhlenbruck and B. B{\"u}chner, Phys. Rev. B {\bf
65}, 174404 (2002).

\bibitem{BCG94}
J. Blasco, M. Castro and J. Garcia, J. Phys.: Condens. Matter {\bf
6}, 5875 (1994).

\bibitem{PBG99}
J. P{\'e}rez, J. Blasco, J. Garc{\'{\i}}a, M. Castro, J.
Stankiewicz, M.C. S{\'a}nchez and R.D. S{\'a}nchez, J. Magn. Magn.
Mater. {\bf 196-197}, 541 (1999).

\bibitem{CZQ05}
C.X. Chen, R.K. Zheng, T. Qian, Y. Liu and X.G. Li, J. Phys. D:
Appl. Phys. {\bf 38}, 807 (2005).

\bibitem{SSS02}
S. Seiro, H.R. Salva, M. Saint-Paul, A.A. Ghilarducci, P. Lejay, P.
Monceau, M. Nunez-Regueiro and A. Sulpice, J. Phys.: Condens. Matter
{\bf 14}, 3973 (2002).

\bibitem{ZSL00}
S. Zvyagin, H. Schwenk, B. L{\"u}thi, K.V. Kamenev, G. Balakrishnan,
D. McK. Paul, V.I. Kamenev and Yu.G. Pashkevic, Phys. Rev. B {\bf
62}, 6104 (2000).

\bibitem{ZGD00}
J.-S. Zhou, J.B. Goodenough, B. Dabrowski, P.W. Klamut and Z.
Bukowski, Phys. Rev. Lett. {\bf 84}, 526 (2000).

\bibitem{Reh73}
W. Rehwald, Adv. Phys. {\bf 22}, 721 (1973).

\bibitem{BRN92}
A. Bulou, M. Rousseau and J. Nouet, Key Eng. Mater. {\bf 68}, 133
(1992).

\bibitem{CS98}
M.A. Carpenter and E.H.K. Salje, Eur. J. Mineral. {\bf 10}, 693
(1998).

\bibitem{GL03}
T. Goto and B. L{\"u}thi, Adv. Phys. {\bf 52}, 67 (2003).

\bibitem{Lut07}
B. L{\"u}thi, {\it Physical Acoustics in the Solid State}. Springer
Series in Solid-State Sciences.  (Springer, Berlin, 2007).

\bibitem{LWS69}
M. Long, Jr., A.R. Wazzan and R. Stern, Phys. Rev. {\bf 178}, 775
(1969).

\bibitem{KMB94}
U. Kawald, O. Mitze, H. Bach, J. Pelz and G.A. Saunders, J. Phys.:
Condens. Matter {\bf 6}, 9697 (1994).

\bibitem{MHI06}
S. Morita, H. Higaki, I. Ishii, M. Takemura, F. Iga, T. Takabatake,
M. Tsubota and T. Suzuki, Physica B {\bf 383}, 43 (2006).

\bibitem{PLP07}
M. Poirier, F. Laliberte, L. Pinsard-Gaudart and A. Revcolevschi,
Phys. Rev. B {\bf 76}, 174426 (2007).

\bibitem{Mel76}
R.L. Melcher, {\it Physical Acoustics}. ed. by W.P. Mason and R.N.
Thurston, p. 1 (Academic Press, New York, 1976).

\bibitem{33}
F. Cordero, Phys. Rev. B {\bf 47}, 7674 (1993).

\bibitem{KYB09}
H. K{\"o}ppel, D.R. Yarkony and H. Barentzen, {\it The Jahn-Teller
Effect - Fundamentals and Implications for Physics and Chemistry}.
(Springer, Heidelberg, 2009).

\bibitem{LB78}
G. Leibfried and N. Breuer, {\it Point Defects in Metals I}.
(Springer, Berlin,
 1978).

\bibitem{BW68}
O.G. Brandt and C.T. Walker, Phys. Rev. {\bf 170}, 528 (1968).

\bibitem{KK82}
K.I. Kugel and D.I. Khomskii, Sov. Phys. Usp. {\bf 25}, 231 (1982).

\bibitem{MFL08}
M. Medarde, M.T. Fern{\'a}ndez-D\'iaz and Ph. Lacorre, Phys. Rev. B
{\bf 78}, 212101 (2008).

\bibitem{BA04}
N. Binggeli and M. Altarelli, Phys. Rev. B {\bf 70}, 085117 (2004).

\end{references}
\end{document}